\documentclass[prb,twocolumn,showpacs,preprintnumbers,amsmath,amssymb]{revtex4-1}
\usepackage{graphicx}

\begin{document}

\title{Effects of Zn substitution on the electronic structure of BaFe$_2$As$_2$ revealed by angle-resolved photoemission spectroscopy}

\author{S.~Ideta$^{1}$, T.~Yoshida$^{2,6}$, M.~Nakajima$^{5,6}$, W.~Malaeb$^{3}$, T.~Shimojima$^{1}$, K.~Ishizaka$^{1}$,  A.~Fujimori$^{2,6}$, H.~Kimigashira$^{4}$, K.~Ono$^{4}$, K.~Kihou$^{5,6}$, Y.~Tomioka$^{5,6}$, C.~H.~Lee$^{5,6}$, A.~Iyo$^{5,6}$, H.~Eisaki$^{5,6}$, T.~Ito$^{5,6}$, and S.~Uchida$^{2,6}$}

\affiliation{$^1$ Department of Applied Physics, University of Tokyo, Bunkyo-ku, Tokyo 113-8656, Japan\\
$^2$ Department of Physics, University of Tokyo, Bunkyo-ku, Tokyo 113-0033, Japan\\
$^3$ Institute for Solid State Physics (ISSP), University of Tokyo, Kashiwa-no-ha, Kashiwa, Chiba 277-8581, Japan\\
$^4$ KEK, Photon Factory, Tsukuba, Ibaraki 305-0801, Japan\\
$^5$ National Institute of Advanced Industrial Science and Technology, Tsukuba 305-8568, Japan\\
$^6$ JST, Transformative Research-Project on Iron Pnictides (TRIP), Chiyoda, Tokyo 102-0075, Japan}
\date{\today}%
\begin{abstract}
In Fe-based superconductors, electron doping is often realized by the substitution of transition-metal atoms for Fe. In order to investigate how the electronic structure of the parent compound is influenced by Zn substitution, which supplies nominally four extra electrons per substituted atom but is expected to induce the strongest impurity potential among the transition-metal atoms, we have performed an angle-resolved photoemission spectroscopy measurements on Ba(Fe$_{1-x}$Zn$_x$)$_2$As$_2$ (Zn-122). In Zn-122, the temperature dependence of the resistivity shows a kink around $T\sim$ 135 K, indicating antiferromagnetic order below the N$\rm{\acute{e}}$el temperature of $T_{\rm{N}}\sim$ 135 K. In fact, folded Fermi surfaces (FSs) similar to those of the parent compound have been observed below $T_{\rm{N}}$. The hole and electron FS volumes are, therefore, different from those expected from the rigid-band model. The results can be understood if all the extra electrons occupy the Zn 3$d$ state $\sim$10 eV below the Fermi level and do not participate in the formation of the FSs.

\end{abstract}

\maketitle

In iron-based high-$T_c$ superconductors (Fe-SCs), electron doping can be made by partial substitution of transition-metal atoms for Fe in the antiferromagnetic (AFM) parent compounds as in Ba(Fe$_{1-x}$Co$_x$)$_2$As$_2$ (Co-122) and Ba(Fe$_{1-x}$Ni$_x$)$_2$As$_2$ (Ni-122), which show relatively high superconducting (SC) transition temperatures of $T_c$ = 25 and 20 K, respectively\cite{Sefat, Canfield, Ni, Olariu}. This is a surprising phenomenon, considering the fact that in the case of cuprate superconductors, transition-metal impurities such as Co, Ni, and Zn works as strong scatterers and quickly kill the superconductivity \cite{HAlloul, JMTarascon, JLTallon, YKKuo}. Therefore, the role of the substituted transition-metal atom in the electronic structure has been one of the most important but difficult issues in the Fe-SCs.  

\begin{figure}[t]
\includegraphics[width=7cm]{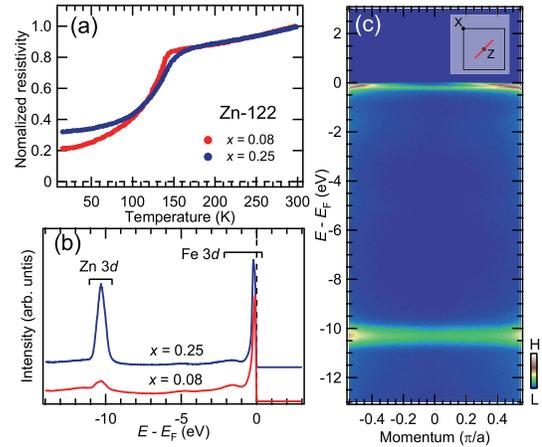}
\caption{(Color online) (Color online) (a): In-plane resistivity normalized at $T$ = 300 K is plotted as a function of temperature for Ba(Fe$_{1-x}$Zn$_x$)$_2$As$_2$ (Zn-122) with $x$ = 0.08 and 0.25. (b): Valence-band spectrum of Zn-122 with $x$ = 0.08 and 0.25 taken at $h\nu$ = 60 eV. (c): ARPES intensity plots of Zn-122 with $x$ = 0.25 around the Z point. A red line of the inset shows a momentum space of the $E$-$k$ plot of (c). }
\label{Fig1}
\end{figure}
\begin{figure*}[t]
\includegraphics[width=17cm]{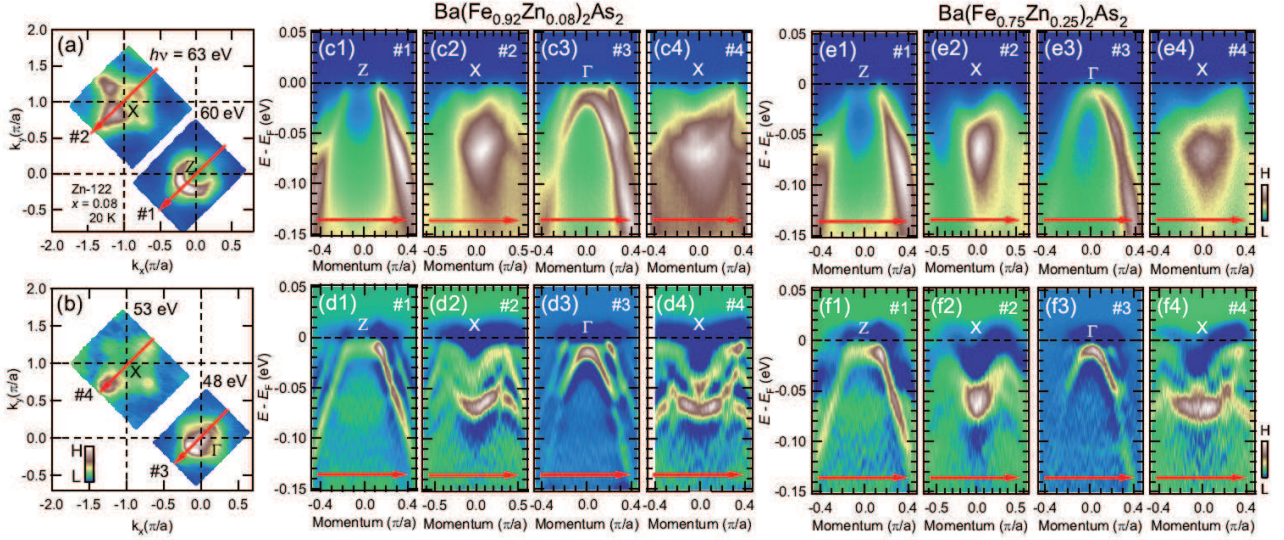}
\caption{(Color online) Fermi surfaces (FSs) and band dispersions of Zn-122 observed by ARPES. (a),(b): ARPES intensity mapping plotted in the two-dimensional $k_x$-$k_y$ plane taken at several photon energies. ARPES intensities have been integrated within $\pm$5 meV from the Fermi level ($E_{\rm{F}}$). (c1)-(f4): Band dispersions of $x$ = 0.08 and 0.25 along cuts $\#$1 to $\#$4 in (a) and (b), and the second-derivative of the energy-momentum plots for the same cuts. Red arrows correspond to the red arrows of (a) and (b).}
\label{Fig2}
\end{figure*}

Based on super-cell calculation within the density functional theory (DFT), Wadati $et\ al$. have studied the spatial distribution of doped electrons in the FeAs layer and found that the doped electrons are located at the Co, Ni, Cu, and Zn sites \cite{Wadati, Bittar}. On the other hand, another super-cell calculation by Konbu $et\ al$. \cite{Konbu} has shown a rigid-band-like shift of the chemical potential and concomitant evolution of FSs with Co and Ni substitution. Our recent angle-resolved photoemission spectroscopy (ARPES) measurement has revealed that deviation from the rigid-band model gradually develops in going from Co-, Ni-, to Cu doped (Cu-122) BaFe$_2$As$_2$, which we attribute to the effect of impurity potential of the substitutional transition-metal atoms \cite{SIdeta_TFeAs}. For Zn substitution, the doped electron number is expected to be four times as large as that of Co doping, and at the same time the impurity potential is expected to be stronger than Co, Ni, and Cu. In fact, the Zn 3$d$ state is located $\sim$8 eV below the Fermi level ($E_{\rm{F}}$) in DFT calculations \cite{Wadati, Berlijn}. The effect of Zn-doping on the $T_c$ of Fe-SC's has been controversial. According to Zn-doping studies on polycrystalline Ba$_{0.5}$K$_{0.5}$Fe$_2$As$_2$ (K-122) \cite{PCheng}, the $T_c$ does not change with Zn doping. For LaFeAsO$_{1-x}$F$_x$ (F-La1111) \cite{YLi}, on the other hand, the $T_c$ even increases with Zn doping in the underdoped region, while it remains unchanged in the optimally doped regions. However, recent studies on single-crystalline K-122 and polycrystalline F-La1111 over a broad electron-doping range have shown that the $T_c$ is strongly suppressed with Zn substitution \cite{YFGuo, JLi1, JLi2}.

In this work, we have investigated the electronic structure of Ba(Fe$_{1-x}$Zn$_x$)$_2$As$_2$ (Zn-122) to study the effect of Zn doping in the BaFe$_2$As$_2$ system, focusing on the impurity states and FS volumes. We have found that the electronic structure of Zn-122 is similar to that of the parent compound BaFe$_2$As$_2$, that is, band folding due to the AFM order remains unchanged. We have also found that the hole and electron FS volumes above $T_{\rm{N}}$ are similar to those of the parent compound and, therefore, are strongly deviated from those expected from the rigid-band model. This implies that doped electrons are mostly localized on the Zn 3$d$ level located $\sim$10 eV below $E_{\rm{F}}$. The present result is in line with the \textquotedblleft universal\textquotedblright\ relationship between the $T_{\rm{N}}$ and the total FS volume for the transition-metal-substituted Fe-SCs \cite{SIdeta_TFeAs}.

High-quality single crystals of Ba(Fe$_{1-x}$Zn$_x$)$_2$As$_2$ with $x$ = 0.08 and 0.25 were grown by the self-flux method. The Zn concentration was determined by energy dispersive x-ray (EDX) analysis. ARPES measurements were carried out at beamline 28A of Photon Factory (PF) using linearly-polarized light ranging from $h\nu$ = 34 to 80 eV. A Scienta SES-2002 analyzer was used with the total energy resolution of 15-20 meV. The crystals were cleaved $in\ situ$ at $T$ = 20 K in an ultra-high vacuum of $\sim$5$\times$10$^{-11}$ Torr.

Figure \ref{Fig1}(a) shows the temperature dependence of the in-plane resistivity of Zn-122. It shows a kink at $T\sim$ 135 K very similar to the resistivity of the parent compound BaFe$_2$As$_2$, and indicates that the AFM order persists in Zn-122. Figure \ref{Fig1}(b) shows the angle-integrated photoemission spectra of Zn-122 in the entire valence-band region. Zn 3$d$-derived emission is seen around 10 eV below $E_{\rm{F}}$, and is well separated from the Fe 3$d$ band, consistent with the density of state (DOS) given by the super-cell band-structure calculation \cite{Wadati}. In Fig. \ref{Fig1}(c), an energy-momentum ($E$-$k$) plot taken at $h\nu$ = 60 eV on the Z plane is shown, and non-dispersive bands are observed 10 eV below $E_{\rm{F}}$. This indicates that Zn-122 shows stronger deviation from the rigid-band model than Co-, Ni-, and Cu-122 in the previous report \cite{SIdeta_TFeAs}.

In order to see the effect of the AFM order on the electronic structure near $E_{\rm{F}}$, we have performed FS mapping in the $k_x$-$k_y$ plane at $T$ = 20 K. Figures \ref{Fig2}(a) and \ref{Fig2}(b) show intensity mapping in the $k_x$-$k_y$ planes including the Z and the $\Gamma$ points, respectively. One finds that the FS shapes for the hole and electron FSs around the Brillouin zone (BZ) center and corner are similar to those of the parent compound BaFe$_2$As$_2$ \cite{MYi1, TShimojima}. As shown in Figs. \ref{Fig2}(c1)-2(f4), we show $E$-$k$ plots and corresponding second-derivative $E$-$k$ plots for the hole and electron bands taken at $T$ = 20 K for cuts $\#$1 - $\#$4 in Figs. \ref{Fig2}(a) and \ref{Fig2}(b). The band folding due to the spin-density-wave (SDW) formation is obviously observed for $x$ = 0.08, while the band dispersions of $x$ = 0.25 is smeared out and the band folding is rather difficult to identify, probably due to scattering by the randomly distributed Zn atoms. Here, the hole bands for both samples clearly cross the $E_{\rm{F}}$ even around the $\Gamma$ point \cite{SIdeta_TFeAs, DJSingh, WMalaeb}, implying that electrons are not doped appreciably by Zn substitution.  
 \begin{figure}[ht]
\includegraphics[width=9cm]{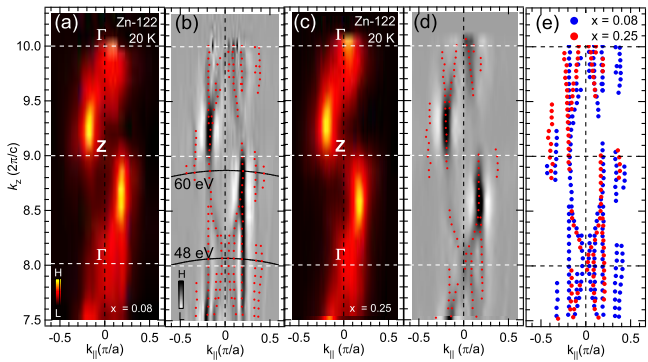}
\caption{(Color online) FS mapping in the $k_{||}$-$k_z$ plane below $T_{\rm{N}}$ $\sim$ 135 K. (a), (c): Hole FSs around the center of the Brillouin zone. (b), (d): Second-derivatives of the intensity map in (a) and (c), respectively. Red dots denote positions of the Fermi momentum ($k_{\rm{F}}$). $k_{\rm{F}}$'s deduced from panels (b) and (d) are shown in panel (e). }
\label{Fig3}
\end{figure} In order to investigate the three-dimensional electronic structure and to estimate the FS volumes quantitatively, FS mapping in the $k_z$-$k_{||}$ plane was performed by changing the photon energy as shown in Fig. \ref{Fig3}. The direction of $k_{||}$ is the same as cuts in Figs. \ref{Fig2}(a) and \ref{Fig2}(b). Here, the intensity asymmetry observed with respect to the $k_{||}$ = 0 line is due to photoemission matrix-element effects. Figures \ref{Fig3}(a) and \ref{Fig3}(c) show the hole FS mapping on the $k_z$-$k_{||}$ plane at $T$ = 20 K (well below $T_{\rm{N}}\sim$ 135 K) for the $x$ = 0.08 and 0.25 samples, respectively, and Figs. \ref{Fig3}(b) and \ref{Fig3}(d) are their second derivatives. Three dimensional hole FSs folded due to AFM order are observed, similar to the previous ARPES and dHvA studies of BaFe$_2$As$_2$ \cite{MYi1, TTerashima, YNakashima, TShimojima}. Figure \ref{Fig3}(e) shows that the shapes of hole FSs for both samples fall on top of each other.

\begin{figure}[t]
\includegraphics[width=7.5cm]{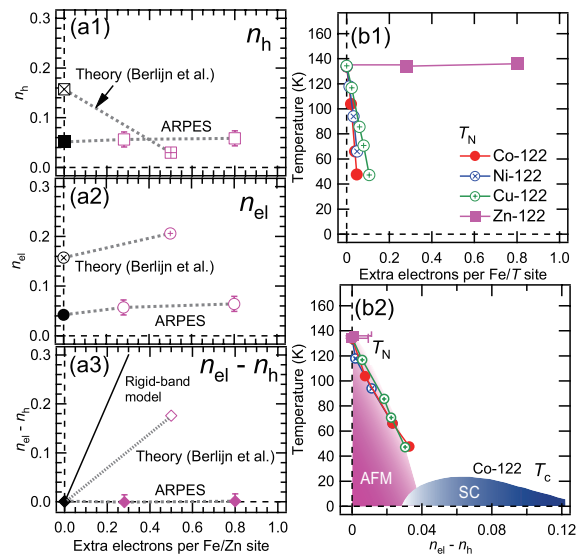}
\caption{(Color online) (a1)-(a3): Hole ($n_{\rm{h}}$), electron ($n_{\rm{el}}$), and total FS volumes ($n_{\rm{el}}-n_{\rm{h}}$) above $T_{\rm{N}}$ plotted as functions of the number of extra electrons. Experimental data obtained from the present study are compared with theory (configuration-averaged super-cell band-structure calculation reported by theory \cite{Berlijn}) and the rigid-band model shown by a solid line in (a3). The data of BaFe$_2$As$_2$ are taken from experimental studies \cite{WMalaeb, Dhaka}. $T_{\rm{N}}$'s plotted as functions of (b1) the nominal extra electron number per Fe/$T$ site and (b2) the total FS volume $n_{\rm{el}}-n_{\rm{h}}$ ($T$ = Co, Ni, Cu, and Zn) \cite{SIdeta_TFeAs}. The $T_c$ dome of Co-122 is also shown for reference.}
\label{Fig4}
\end{figure}

In order to estimate the volumes of the hole and electron FSs in the paramagnetic state, we have performed FS mapping above $T_{\rm{N}}$ (see Supplemental Material\cite{Supplemental}). Estimated hole ($n_{\rm{h}}$), electron ($n_{\rm{el}}$), and total ($n_{\rm{el}}-n_{\rm{h}}$) FS volumes are plotted as functions of the number of extra electrons per Fe/Zn site in Figs. \ref{Fig4}(a1)-\ref{Fig4}(a3), respectively. Here, the number of extra electrons has been calculated using the net Zn concentrations ($x \sim$ 0.07 and $\sim$ 0.21 for $x$ = 0.08 and 0.25 samples, respectively \cite{Supplemental}) which excludes the precipitated phase of Zn metal from the total Zn concentration estimated by the EDX analysis. We have assumed doubly degenerate inner and one outer hole FSs. The FS volumes deduced from the present study are compared with configuration-averaged super-cell band-structure calculation \cite{Berlijn} and the rigid-band model, $n_{\rm{el}}-n_{\rm{h}}$ = $x$, in Fig. \ref{Fig4}. While the calculated $n_{\rm{h}}$ decreases only slightly (by $\sim$0.1) with Zn substitution compared to the added electron number of 0.5, $n_{\rm{h}}$ deduced from ARPES shown an even smaller change. In particular, the $n_{\rm{el}}$ of Zn-122 is almost constant unlike the theoretical prediction. The experimental value of $n_{\rm{el}}-n_{\rm{h}}$ thus does not show appreciable change from the non-doped sample as shown in Fig. \ref{Fig4}(a3), indicating that the rigid-band model completely breaks down in Zn-122. We have plotted the $T_{\rm{N}}$ of Zn-122 as a function of the extra electron number [Fig. \ref{Fig4}(b1)] and of $n_{\rm{el}}-n_{\rm{h}}$ deduced from the present study [Fig. \ref{Fig4}(b2)] together with the other transition-metal-substituted BaFe$_2$As$_2$. While, the $T_{\rm{N}}$ plotted in Fig. \ref{Fig4}(b1) is strongly deviated between Co-, Ni-, Cu-, and Zn-122, the $T_{\rm{N}}$ plotted in Fig. \ref{Fig4}(b2) coincides with each other \cite{SIdeta_TFeAs}, indicating that the $T_{\rm{N}}$ is controlled by the number of doped electrons. Therefore, almost all the doped electrons do not become mobile carriers, and hence, Zn-122 does not show the superconductivity.

\begin{figure}[t]
\includegraphics[width=8.8cm]{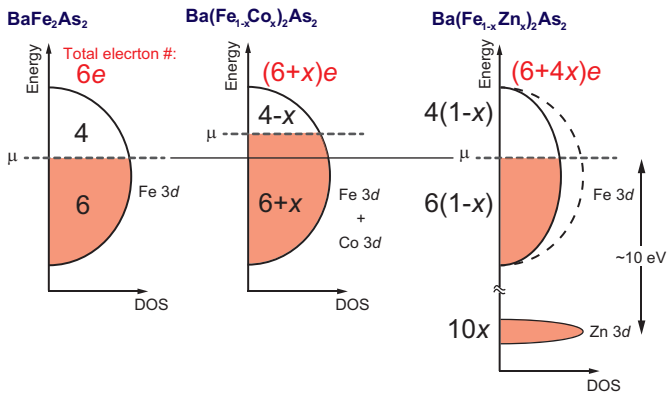}
\caption{(Color online) Schematic illustration of the density of states for BaFe$_2$As$_2$, Ba(Fe$_{1-x}$Co$_x$)$_2$As$_2$ (Co-122), and Ba(Fe$_{1-x}$Zn$_x$)$_2$As$_2$ (Zn-122). The total electron numbers of BaFe$_2$As$_2$, Co-122, and Zn-122 are 6, (6+$x$), and (6+4$x$), per Fe/Co/Zn site, respectively.}
\label{Fig5}
\end{figure}

Now, we discuss how the electronic structure is influenced by the strong impurity potential of the Zn atom. As shown in Fig. \ref{Fig3}, changes in the FSs induced by Zn substitution are very little, and the electronic structure of Zn-122 is similar to that of BaFe$_2$As$_2$. According to tight-binding-model calculations and first-principles band-structure calculations \cite{MWHaverkort}, in the case of a weak impurity potential, the host and impurity atomic orbitals form common energy bands. With increasing magnitude of the impurity potential, the host Fe 3$d$ and the impurity 3$d$ bands are separated, and the impurity states are split off below the host band. Thus, the capacity of the impurity band to accommodate additional electrons increases, and the number of electrons added to the FSs decreases. In order to explain the present ARPES result that the FSs and band structure show almost no change with Zn substitution, schematic DOSs for BaFe$_2$As$_2$, Co-122, and Zn-122 are drawn in Fig. \ref{Fig5}. For BaFe$_2$As$_2$, the Fe 3$d$ bands (hybridized with As 4$p$ bands) are occupied by 6 electrons per Fe. By substitution of Co for Fe, because the potential difference $\leq\ \sim$2 eV \cite{Wadati} is small compared to the total bandwidth $\sim$ 2-3 eV, the hybridized common Fe 3$d$-Co 3$d$ bands are formed. Thus, the chemical potential ($\mu$) is shifted upward to accommodate the 6+$x$ electrons per Fe/Co site. On the other hand, as shown in Fig. \ref{Fig5}, the Zn 3$d$ level in Zn-122 exists at the high binding energy of $\sim$10 eV and the number of electrons doped by Zn substitution 10$x$ occupy the Zn 3$d$ level which can accommodate 10$x$ electrons per Fe/Zn site, while the Fe 3$d$ bands can now accommodate only 10(1-$x$) electrons because of the missing Fe atoms. Since the total electron number in Zn-122 is 6+4$x$, the occupancy of the Fe 3$d$ bands must be 6(1-$x$). Consequently, the ratio of the occupied and unoccupied states of the Fe 3$d$ bands in Zn-122 is 4 : 6, the same as that of BaFe$_2$As$_2$, and hence the chemical potential remains unshifted. Finally, it is interesting to compare the present result with the isovalent-doped Ba(Fe$_{1-x}$Ru$_x$)$_2$As$_2$ (Ru-122), where the chemical potential also does not shift as in the case of Zn-122 but superconductivity appears. According to the previous ARPES study\cite{Dhaka2}, the Ru 4$d$ states are located around -1 eV below $E_{\rm{F}}$ and, therefore, will be merged into the Fe 3$d$ bands. Thus, the Fe 3$d$-Ru 4$d$ bands near $E_{\rm{F}}$ are broadened compared to those of the parent compound, the antiferromagnetic ordering is suppressed, and the superconductivity appears.

In conclusion, we have performed an ARPES study of Ba(Fe$_{1-x}$Zn$_x$)$_2$As$_2$ with $x$ = 0.08 and 0.25 to investigate the effect of Zn doping on the electronic structure. From the three-dimensional shape of the FSs determined by ARPES, we found that the hole and electron FS volumes do not change, that is, a dramatic deviation from the rigid-band model occurs. Since very few electrons are doped into the Fe 3$d$ band, the AFM persists in Zn-122, and we have confirmed the universal relationship between the $T_{\rm{N}}$ and $n_{\rm{el}}-n_{\rm{h}}$ among the transition metal-doped BaFe$_2$As$_2$ systems \cite{SIdeta_TFeAs}.

\section*{ACKNOWLEDGMENTS}

ARPES experiments were carried out at KEK-PF under the approval of the PF Program Advisory Committee (Proposals No. 2012G751, 2012G075, and 2012S2-001). This work was supported by an A3 Foresight Program from the Japan Society for the Promotion of Science, a Grant-in-Aid for Scientific Research on Innovative Area gMaterials Design through Computics: Complex Correlation and Non-Equilibrium Dynamicsh, and a Global COE Program gPhysical Sciences Frontierh, MEXT, Japan. S.I. acknowledges support from the Japan Society for the Promotion of Science for Young Scientists.

\section*{Supplemental Material}
\subsection{Fermi-surface mapping in the $k_x$-$k_y$ plane and band dispersions measured above {$T_N$}}

\begin{figure*}[h]
\includegraphics[width=18cm]{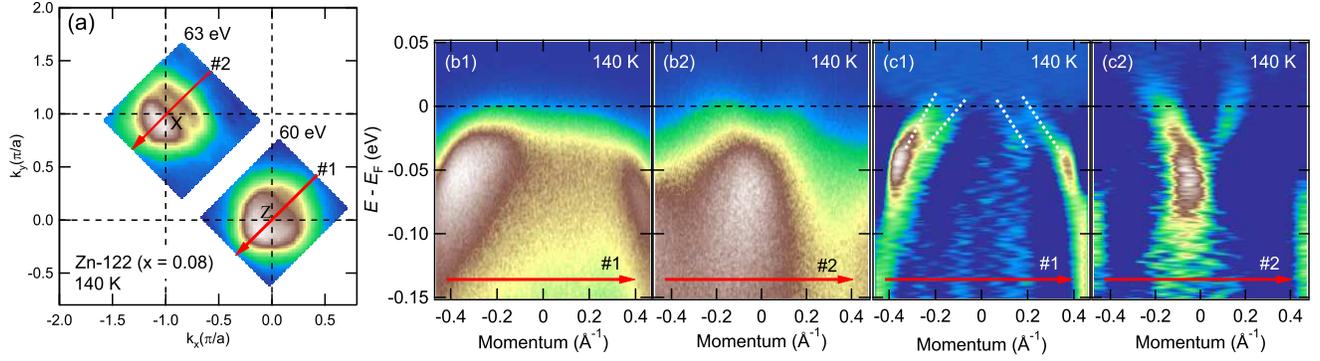}
\caption{(a): Fermi-surface (FS) mapping plotted in the $k_x$-$k_y$ plane of Zn-122 with $x$ = 0.08. (b1), (b2): Energy-momentum plots of the ARPES intensity for cuts $\#1$ and $\#2$ in panel (a). (c1), (c2): Second derivatives of the MDCfs in panels (b1) and (b2). White dotted lines on panel (c1) are guide to the eye.}
\label{FigS1}
\end{figure*}

Fermi-surface (FS) mapping and hole and electron band dispersions taken above $T_N$ are shown in Fig. \ref{FigS1}. The reconstruction of the FSs due to antiferromagnetic order is absent, and paramagnetic hole and electron FSs are observed. Hole and electron bands are clearly observed as shown in Figs. \ref{FigS1}(c1) and \ref{FigS1}(c2), respectively.

\subsection{Fermi-surface mapping in the $k_z$-$k_{||}$ plane measured above $T_N$}

\begin{figure*}[h]
\includegraphics[width=13cm]{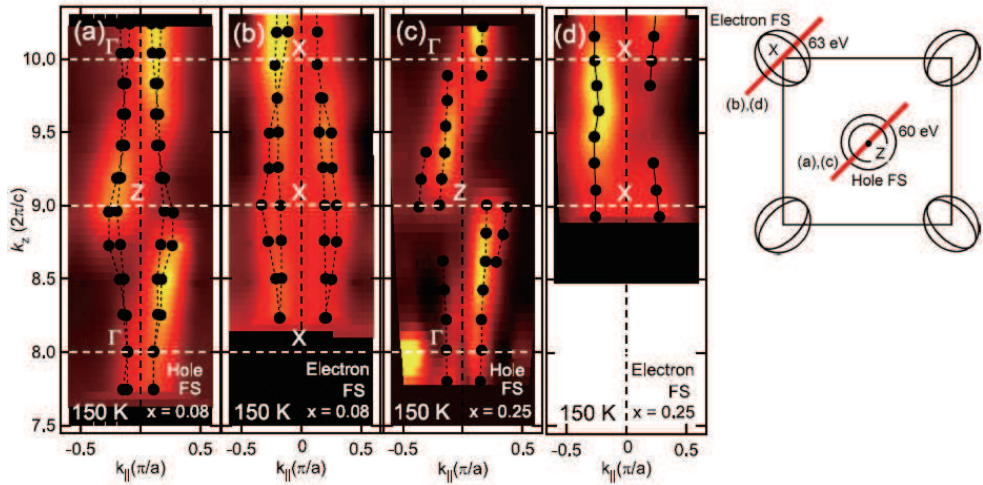}
\caption{FS mapping in the $k_z$-$k_{||}$ plane above $T_N$. (a), (c): Hole FS mapping. (c), (d): Electron FS mapping. The right panel shows a schematic Brillouin zone (BZ) in the $k_x$-$k_y$ plane. Hole and electron FSs are also shown at the BZ center and corner, respectively. The momentum direction of panels (a), (c) and (b), (d) are shown by red bars.}
\label{FigS2}
\end{figure*}

FS mapping in the $kz$-$k_{||}$ plane above $T_N$ has been measured at $T$ = 150 K for $h\nu$ = 34-84 eV with linearly polarized light for Zn-122 in order to estimate the FS volumes (see text) as shown in Fig. \ref{FigS2}. We have found that the hole and electron FSs show warping, while the shape of the electron FS is almost straight along the $k_z$ direction. Two hole FSs are resolved for the $x$ = 0.08 and 0.25 samples. For the electron FSs, we have found two sheets for the $x$ = 0.08 sample, while one cannot resolve the two FSs separately for $x$ = 0.25, that is, the electron FS of $x$ = 0.25 are nearly degenerate. Therefore, we have estimated the electron FS volume by assuming two degenerate FSs.

\subsection{Zn 3$d$ level photoemission spectra}

\begin{figure*}[h]
\includegraphics[width=10cm]{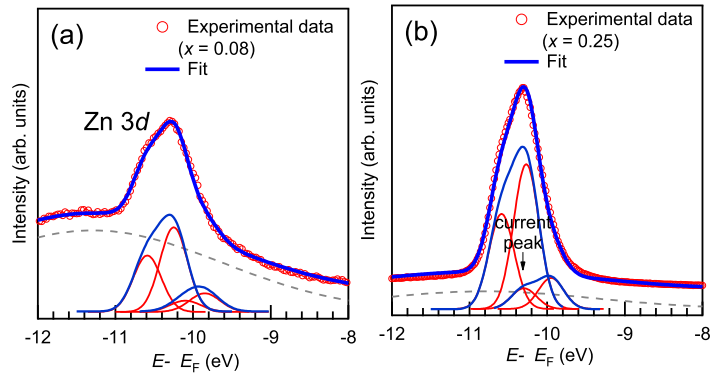}
\caption{Angle-integrated photoemission spectra of the Zn 3$d$ level for $x$ = 0.08 (a) and 0.25 (b).}
\label{FigS3}
\end{figure*}

Since the solubility of Zn in Zn-122 is low, metallic Zn may coexist with Zn substituting Fe. If such an impurity phase of the metallic Zn exists, the spectra of the Zn 3$d$ states would show multiple components. Figure S3 shows the Zn 3$d$ photoemission spectra of Zn-122 with $x$ = 0.08 and 0.25. The Zn 3$d$ states have narrow bandwidth and can be regarded as core levels with the spin-orbit doublet of $d_{5/2}$ and $d_{3/2}$. Therefore, we have fitted each Zn 3$d$ spectrum to a superposition of two spin-orbit doublets. Each spin-orbit doublet consists of two set of two Gaussians with the intensity ratio $d_{5/2}$:$d_{3/2}$ = 3:2 (red curves in Fig. \ref{FigS3}). The two spin-orbit doublets are separated due to a chemical shift between the Zn-122 phase and the impurity phase. The intensity ratio of the two doublets was found to be 5:1 for the $x$ = 0.08 sample and 7:1 for the $x$ = 0.25 sample, that is, the net Zn concentration without the impurity phase is deduced to be $\sim$ 0.07 and $\sim$ 0.2 for the $x$ = 0.08 and $x$ = 0.25 samples, respectively. We have used the Zn concentrations in the Zn-122 phase thus deduced rather than the nominal concentrations throughout this paper.


\begin{thebibliography}{99}

\bibitem{Sefat} A. S. Sefat, R. Jin, M. A. McGuire, B. C. Sales, D. J. Singh, and D. Mandrus, Phys. Rev. Lett. \textbf{101}, 117004 (2008).
\bibitem{Canfield} P. C. Canfield, S. L. Bud'ko, Ni Ni, J. Q. Yan, and A. Kracher, Phys. Rev. B \textbf{80}, 060501(R) (2009).


\bibitem{Ni}N. Ni, A. Thaler, J. Q. Yan, A. Kracher, E. Colombier, S. L. Bud'ko, P. C. Canfield, and S.T. Hannahs, Phys. Rev. B \textbf{82}, 024519 (2010).


\bibitem{Olariu} A. Olariu, F. Rullier-Albenque, D. Colson, and A. Forget, Phys. Rev. B \textbf{83}, 054518 (2011).

\bibitem{HAlloul} H. Alloul, J. Bobroff, M. Gabay, and P. J. Hirschfeld, Rev. Mod. Phys. \textbf{81}, 45 (2009).

\bibitem{JMTarascon} J. M. Tarascon, E. Wang, S. Kivelson, B. G. Bagley, G. W. Hull, and R. Ramesh, Phys. Rev. B \textbf{42}, 218 (1990).

\bibitem{JLTallon} J. L. Tallon, Phys. Rev. B \textbf{58}, 5956 (1998).

\bibitem{YKKuo} Y. K. Kuo, C. W. Schneider, M. J. Skove, M. V. Nevitt, G. X.
Tessema, and J. J. McGee, Phys. Rev. B \textbf{56}, 6201 (1997).

\bibitem{Wadati} H. Wadati, I. Elfimov, and G. A. Sawatzky, Phys. Rev. Lett. \textbf{105}, 157004 (2010).

\bibitem{Bittar} E. M. Bittar, C. Adriano, T. M. Garitezi, P. F. S. Rosa, L. Mendon\c{c}a-Ferreira, F. Garcia, G. de M. Azevedo, P. G. Pagliuso, and E. Granado, Phys. Rev. Lett. \textbf{107}, 267402 (2011).

\bibitem{Konbu} S. Konbu, K. Nakamura, H. Ikeda, and R. Arita, J. Phys. Soc. Jpn. 80, 123701 (2011).


\bibitem{SIdeta_TFeAs} S. Ideta, T. Yoshida, I. Nishi, A. Fujimori, Y. Kotani, K. Ono, Y. Nakashima, S. Yamaichi, T. Sasagawa, M. Nakajima, K. Kihou, Y. Tomioka,  C. H. Lee, A. Iyo, H. Eisaki, T. Ito, S. Uchida, and R. Arita, Phys. Rev. Lett. \textbf{110}, 107007 (2013).

\bibitem{Berlijn} T. Berlijn, C.-H. Lin, W. Garber, and W. Ku, Phys. Rev. Lett. \textbf{108}, 207003 (2012).

\bibitem{PCheng}P. Cheng, B. Shen, J. Hu, and H-H. Wen, Phys. Rev. B \textbf{81}, 174529 (2010).
 
\bibitem{YLi} Y. Li, J. Tong, Q. Tao, C. Feng, G. Cao, Z. Xu, W. Chen, and
F. Zhang, New J. Phys. \textbf{12}, 083008 (2010).

\bibitem{YFGuo} Y. F. Guo, Y. G. Shi, S. Yu, A. A. Belik, Y. Matsushita, M. Tanaka, Y. Katsuya, K. Kobayashi, I. Nowik, I. Felner, V. P. S. Awana, K. Yamaura, and E. Takayama-Muromachi, Phys. Rev. B \textbf{82}, 054506 (2010).

\bibitem{JLi1} J. Li, Y. Guo, S. Zhang, S. Yu, Y. Tsujimoto, H. Kontani,
K. Yamaura, and E. Takayama-Muromachi, Phys. Rev. B \textbf{84}, 020513(R) (2011).

\bibitem{JLi2} J. Li, Y. F. Guo, S. B. Zhang, J. Yuan, Y. Tsujimoto, X. Wang, C. I. Sathish, Y. Sun, S. Yu, W. Yi, K. Yamaura, E. Takayama-Muromachiu, Y. Shirako, M. Akaogi, and H. Kontani, Phys. Rev. B \textbf{85}, 214509 (2012).

\bibitem{MYi1} M. Yi, D. H. Lu, J. G. Analytis, J.-H. Chu, S.-K. Mo, R.-H. He, M. Hashimoto, R. G. Moore, I. I. Mazin, D. J. Singh, Z. Hussain, I. R. Fisher, and Z.-X. Shen, Phys. Rev. B \textbf{80}, 174510 (2009).

\bibitem{DJSingh}D. J. Singh, Phys. Rev. B \textbf{78}, 094511 (2008).

\bibitem{WMalaeb} W. Malabe, T. Yoshida, A. Fujimori, M. Kubota, K. Ono, K. Kihou, P. M. Shirage, H. Kito, A. Iyo, H. Eisaki, Y. Nnakajima, T. Tamegai, and R. Ariata, J. Phys. Soc. Jpn. \textbf{78}, 123706 (2009).

\bibitem{TTerashima} T. Terashima, N. Kurita, M. Tomita, K. Kihou, C.-H. Lee, Y. Tomioka, T. Ito, A. Iyo, H. Eisaki, T. Liang, M. Nakajima, S. Ishida, S.I. Uchida, H. Harima, and S. Uji, Phys. Rev. Lett. \textbf{107}, 176402 (2011).

\bibitem{YNakashima} Y. Nakashima, A. Ino, S. Nagato, H. Anzai, H. Iwasawa, Y. Utsumi, H. Sato, M. Arita, H. Namatame, M. Taniguchi, T. Oguchi, Y. Aiura, I. Hase, K. Kihou, C. H. Lee, A. Iyo, H. Eisaki, Solid St. Commun. \textbf{157}, 16 (2013).

\bibitem{TShimojima} T. Shimojima, K. Ishizaka, Y. Ishida, N. Katayama, K. Ohgushi, T. Kiss, M. Okawa, T. Togashi, X.-Y. Wang, C.-T. Chen, S. Watanabe, R. Kadota, T. Oguchi, A. Chainani, and S. Shin, Phys. Rev. Lett. \textbf{104}, 057002 (2010).


\bibitem{Supplemental} See Supplemental Material at [URL will be inserted by publisher] for details about the ARPES spectra in the normal state and how to estimate the net Zn concentration.

\bibitem{MWHaverkort} M. W. Haverkort, I. S. Elfimov, and G. A. Sawatzky, arXiv:1109.4036.

\bibitem{Dhaka} R. S. Dhaka, S. E. Hahn, E. Razzoli, R. Jiang, M. Shi, B. N. Harmon, A. Thaler, S. L. Bud'ko, P. C. Canfield, and A. Kaminski, Phys. Rev. Lett. \textbf{110}, 067002 (2013).

\bibitem{Dhaka2} R. S. Dhaka, Chang Liu, R. M. Fernandes, Rui Jiang, C. P. Strehlow, Takeshi Kondo, A. Thaler, J$\ddot{o}$rg Schmalian, S. L. Bud'ko, P. C. Canfield, and Adam Kaminski, Phys. Rev. Lett. \textbf{107}, 267002 (2011).



\end{thebibliography}
\end{document}